\documentclass[graybox]{svmult}

\usepackage{type1cm}        
\usepackage{makeidx}         
\usepackage{graphicx}        
\usepackage{multicol}        
\usepackage[bottom]{footmisc}

\usepackage{newtxtext}       %
\usepackage[varvw]{newtxmath}       

\makeindex             

\begin{document}

\title*{Deep Learning 21cm Lightcones in 3D}
\author{Caroline Heneka}
\institute{Caroline Heneka, Institute for Theoretical Physics, Heidelberg University, Philosophenweg 16, 69120 Heidelberg, Germany, \\ \email{heneka@thphys.uni-heidelberg.de} 
}
%
%
\maketitle

\abstract*{Interferometric measurements of the 21cm signal are a prime example of the data-driven era in astrophysics we are entering with current and upcoming experiments. We showcase the use of deep networks that are tailored for the structure of 3D tomographic 21cm light-cones to firstly detect and characterise HI sources and to secondly directly infer global astrophysical and cosmological model parameters. We compare different architectures and highlight how 3D CNN architectures that mirror the data structure are the best-performing model. 
}

\abstract{Interferometric measurements of the 21cm signal are a prime example of the data-driven era in astrophysics we are entering with current and upcoming experiments. We showcase the use of deep networks that are tailored for the structure of 3D tomographic 21cm light-cones to firstly detect and characterise HI sources and to secondly directly infer global astrophysical and cosmological model parameters. We compare different architectures and highlight how 3D CNN architectures that mirror the data structure are the best-performing model. 
}

\section{21cm imaging in 3D to track the history of the Universe}
\label{sec:intro}
The 21cm signal originates from the forbidden spin-flip transition of neutral hydrogen. As hydrogen is the most abundant element in the Universe it is thus an optimal tracer of structure formation across cosmic time. At high redshift we can map heating due to the very first galaxies during cosmic dawn (CD), followed by the Epoch of Reionisation (EoR) when  neutral hydrogen is ionised again. But also late-time structure formation and galaxy evolution is traced by the 21cm signal via neutral hydrogen within sources. Amazingly, the Square Kilometre Array (SKA, https://www.skao.int/) will achieve 3D tomography and thus produce as data lightcones of 2D spatial maps and finely gridded frequency information. SKA-MID at frequencies of 350 MHz to 15 GHz detects HI sources post-reionisation, SKA-LOW targets at lower frequencies the high-redshift Universe. Due to up to TB/s of data inflowing from SKA, and a highly non-Gaussian signal, deep learning is suitable for the challenge to reduce and analyse this data. Important goals are both source detection for clustering studies and inference of global astrophysical and cosmological parameters. We will highlight here our findings regarding 21cm tomography and optimal deep learning models for both these tasks.

\section{Low-redshift in 3D: 21cm source detection}
\label{sec:SKAchallenge}
A 3D data cube as expected for the SKA was presented as a 1TB cube for the second SKA science data challenge (SDC2, https://sdc2.astronomers.skatelescope.org) for detection and characterisation of >100.000 HI sources (galaxies) centred at a redshift of $z = 0.235 - 0.495$. [1] Together with the full cube a 1GB cube with ground truth information was provided for training. 

Our team 'coin' optimised a couple of modern machine learning algorithms for detection (segmentation) of sources in the SDC2 HI cube and characterisation of source properties. These were integrated HI flux, line width in frequency, HI spatial size and inclination as well as position angle. In addition, a 'classical' detection algorithm based on wavelet filtering for de-noising and segmentation was compared to as a baseline. We considered the following architectures for object detection: 2D and 3D U-Nets,$\,$[2] variants of R-CNN (region-proposal convolutional neural network [3]) and an inception-style network [4] that mimics filtering with wavelets. Main challenges we identified, where data volume is not a problem for a trained network, are 1) low signal-to-noise, 2) small spatial size (few pix), 3) unknown systematics (f. ex. residual beam effects, radio frequency interference, RFIs). 

\begin{figure}[h!]
\sidecaption
\includegraphics[scale=.65]{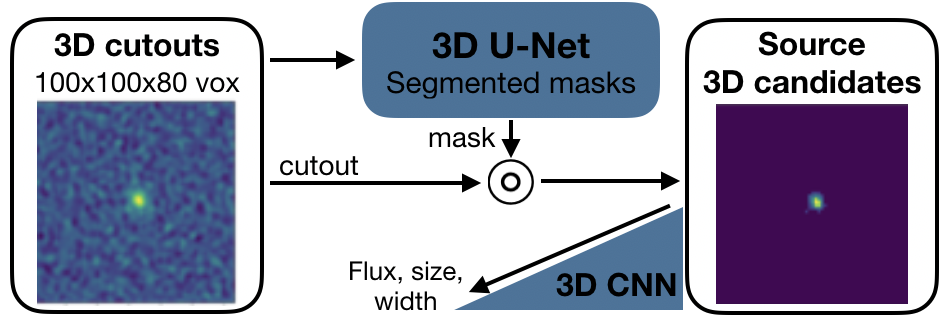}
%
%
\caption{Schematic of our 3D source detection and reconstruction, and HI source property derivation.}
\label{fig:SDC2}       
\end{figure}

We found that a shallow segmentation U-Net architecture generalised to 3D performed best. Trained on 3D cubic patches that each contain a source this architecture is able to provide a 3D reconstruction of HI sources without any pre-processing. Subsequent shallow 3D CNN regression models applied to the high-quality reconstructed images allowed for low-bias and high precision recovery of flux and HI size, almost independently of source properties such as brightness. Without pre-processing rates of false positives reached moderate levels ($\sim$50$\%$), which can be further mitigated  by noise modelling and pre-processing, training on an improved (higher signal-to-noise) fiducial training set, and an extra network-based decision step. Fig.~\ref{fig:SDC2} showcases such a reconstruction; key points to emphasise are:
\begin{itemize}
\item 3D networks perform best for 3D segmentation of HI sources,
\item 3D U-Nets provide high-fidelity 3D reconstructions,
\item which allows for recovery of properties such as HI flux and size \textit{almost independent of source brightness}.
\end{itemize}

\section{High-redshift in 3D: 21cm lightcone inference}
\label{sec:PieNet}
Moving from low-redshift to high-redshift measurements of the 21cm signal, we showcase direct likelihood-free inference from 3D tomography during CD and EoR. To infer a set of astrophysical and cosmological parameters,\footnote{Which here are: Typical X-ray brightness $L_\mathrm{X}$, energy threshold for self-absorption in galaxies $E_0$, ionising efficiency $\zeta$, minimum virial temperature for efficient star formation $T_\mathrm{vir}$, matter density $\Omega_\mathrm{m}$ and warm dark matter mass $m_\mathrm{WDM}$.} we have different options to treat 3D data optimally with network architectures:
\begin{itemize}
\item slicing for set of spatial images $\rightarrow$ 2D CNN, or e.g. ResNet with skip connections,\footnote{ResNet = Residual Network, uses skip connections between residual blocks.}
\item a time-series (frequency-series) of co-eval (spatial) images $\rightarrow$ LSTM network,\footnote{LSTM = long-short term memory network, efficiently memorises different parts of a series for feedback connections. LSTM have been successfully employed for video and audio analysis. [5]}
\item full 3D convolution $\rightarrow$ 3D CNN, moving from 2D to full 3D convolution.
\end{itemize}
We again found a shallow 3D CNN architecture as the best-performing model, in terms of low scatter and bias on inferred parameters, as well as fast and good convergence during training (low loss while avoiding overfitting). The 3D-21cmPIE-Net [4] consists of only four 3D-CNN layers, followed by a GAP (global average pooling) layer against overfitting and three densely connected layers to regress parameters.\footnote{3D 21cm Parameter InfErence Network: https://github.com/stef-neu/3D-21cmPIE-Net} The model was trained on a database of $\sim$5000 lightcones of 140x140x2350 pix at 1.4$\,$Mpc resolution. The training took only $\sim$30 epochs (20min/epoch with a moderate K80 GPU).  

The 3D-21cmPIE-Net was able to jointly constrain cosmology, CD and EoR astrophysics without degeneracies and with small scatter and low bias for most parameters. In detail, $L_\mathrm{X}$ was estimated with 1$\sigma$ errors $<1\%$, $\Omega_\mathrm{m}$ and $T_\mathrm{vir}$ of few $\%$, $\zeta$ down to few $\%$ depending on centre value, $E_0$ at the $10\%$ level and $m_\mathrm{WDM}$ down to roughly $10\%$ (considerably higher for $m_\mathrm{WDM}\gtrsim 2\,$keV). Interestingly, for all parameters except $m_\mathrm{WDM}$ the network's inference was insensitive to increasing levels of foregrounds (instrumental noise, 21cm foreground wedge, horizon and beam effects). Even transfer learning, i.e. training on simulations and evaluating on mock data with noise, and vice versa, yielded low scatter and bias especially for the X-ray luminosity $L_\mathrm{X}$ (and $<1\%$ bias across parameter range), but also robust inference of ionising efficiency $\zeta$. This gives interesting insights on the effect of wrongly simulating noise levels in training data. For example for $\Omega_\mathrm{m}$ training on noiseless simulations and evaluating on mocks yields increased scatter, while the other way round creates bias towards low $\Omega_\mathrm{m}$ values. Thus our 3D network can:
\begin{itemize}
\item efficiently infer both cosmology and astrophysics from full 21cm lightcones,
\item provide low scatter and bias estimates of a multi-parameter set,
\item estimate parameters insensitive to increased noise levels (except for $m_\mathrm{WDM}$),
\item robustly transfer learn simulations $\leftrightarrow$ mock observations for some parameters. 
\end{itemize}

\begin{figure}
\sidecaption
\includegraphics[scale=.24]{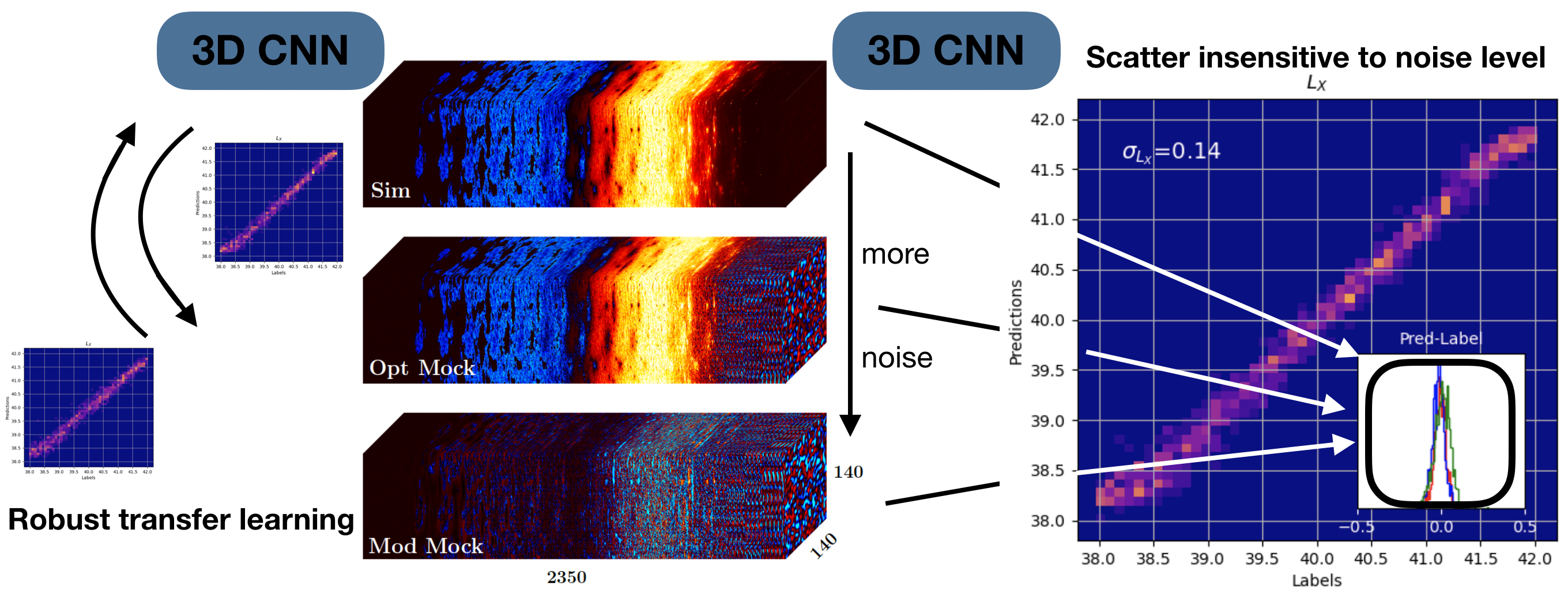}
%
%
\caption{Scheme of key findings for 3D CNN parameter inference from 21cm lightcones [6].}
\label{fig:PieNet}       
\end{figure}

\section{Discussion}
The main message is that \textbf{3D networks for 3D data work best}. This was tested for both low-redshift and high-redshift 3D data cubes of the 21cm signal that traces neutral hydrogen across cosmic times. For low-redshift simulated data in preparation of SKA-MID an HI source detection and characterisation task was performed best by a 3D U-Net and CNN variants as compared to 2D U-Nets, R-CNN and inception models. The 3D U-Net was able to provide 3D reconstructions of sources without preprocessing suitable for estimation of source properties such as HI flux, independent of brightness. For high-redshift simulated data in preparation of SKA-LOW during CD and EoR, an inference task of astrophysical and cosmological parameters was performed best by a 3D CNN as compared to 2D CNN, ResNet, and LSTM models. Remarkably, the quality of the estimate was mostly insensitive to noise (see Fig. 2).

\begin{acknowledgement}
A special thanks and acknowledgement go to S. Neutsch as the main developer of the 3D-21cmPIE-Net, as well as to the team 'coin' of the SKA SDC2 and developers of our machine learning models, M. D. Veneri (INAF), B. Fraga (CBPF), A. Soroka (MSU), F. Gubanov (MSU), C. de Bom (CBPF), A. Meshcheryakov (IKI), and M. Br\"uggen (UHH).
\end{acknowledgement}

%
%
%

\end{document}